# Cyber Defense as a Complex Adaptive System:
## A model-based approach to strategic policy design

Michael D. Norman
mnorman@mitre.org
Center for Connected Government
The MITRE Corporation

Matthew T. K. Koehler
mkoehler@mitre.org
Center for Connected Government
The MITRE Corporation

*Abstract* — In a world of ever-increasing systems interdependence, effective cybersecurity policy design seems to be one of the most critically understudied elements of our national security strategy. Enterprise cyber technologies are often implemented without much regard to the interactions that occur between humans and the new technology. Furthermore, the interactions that occur between individuals can often have an impact on the newly employed technology as well. Without a rigorous, evidence-based approach to ground an employment strategy and elucidate the emergent organizational needs that will come with the fielding of new cyber capabilities, one is left to speculate on the impact that novel technologies will have on the aggregate functioning of the enterprise. In this paper, we will explore a scenario in which a hypothetical government agency applies a complexity science perspective, supported by agent-based modeling, to more fully understand the impacts of strategic policy decisions. We present a model to explore the socio-technical dynamics of these systems, discuss lessons using this platform, and suggest further research and development.

# Introduction and Background

This paper deals with cyber defense decision making at a hypothetical large government agency. As evidenced by the recent creation of the U.S. Senate Cybersecurity Subcommittee (http://www.armed-services.senate.gov/about/subcommittees), the existence of US Cyber Command (http://www.stratcom.mil/Media/Factsheets/Factsheet-View/Article/960492/us-cyber-command-uscybercom/) and specific priorities within the Federal Bureau of Investigation (https://www.fbi.gov/investigate/cyber), and Department of Homeland Security (https://www.dhs.gov/topic/combating-cyber-crime), the US Government is critically interested in designing effective cybersecurity systems and policies. However effective these high-level efforts may be, the first line of defense in cyberspace is the individual system connected to the internet: the device coupled with the human directing it. Collectively, these individual systems are termed "end hosts" and the specific hardware involved include computers, phones, and other web-enabled devices. These systems are typically protected with passwords, private keys, firewalls, and traditional antivirus (AV) or antimalware software. Traditional AV solutions use a 'black listing' methodology (Beuhring 2014), which blocks *already known* threats from being installed and/or executing. This type of methodology operates exclusively in a reactive mode; opening up the enterprise to 'zero-day,' or recently discovered, exploit vulnerability. The protection of end hosts and their local networks is known as cyber defense. The following discussion will focus on the AV solution space and prevention of unwanted code installation/execution.

# Application Whitelisting

Unfortunately, traditional virus scanning is based on previously defined malware and is, by definition, reactive. This means that end hosts will always be vulnerable to previously unknown exploits which can represent significant risk to a large enterprise with many end hosts. However, there is a newly created method for the protection of end hosts specifically designed to deal with this problem and will form the focus of this work. This class of strategy, which is the subject of ongoing acquisition program within our hypothetical government agency, is known as Application Whitelisting (AW). AW allows only previously approved binaries to execute on a given end host and blocks everything else from executing by default (Pareek, 2012). This approach is in stark contrast to traditional AV type solutions, which only block execution of binaries containing known malicious signatures.

On its face, AW appears to solve the virus and malware problem. Unfortunately, our hypothetical government agency is quite large with a diverse set of missions (similar to many other large enterprises). This means there are many end hosts in the agency with different missions requiring various software packages which may change over time. Luckily, there are two basic ways to employ AW:

## Deployment Philosophy 1: "Bottom-up"

This deployment philosophy takes the approach of whitelisting *all* existing applications on an end host at time of deployment. In effect, this creates a custom whitelist for each host. Since these whitelists are defined by the hosts themselves, we refer to this type of deployment methodology as "bottom-up". A process is then instituted for requesting the addition of new applications.

## Deployment Philosophy 2: "Top-down"

This deployment philosophy takes the approach of creating a single whitelist of *all known good* applications only. This single, homogenous whitelist is then distributed to each host in the enterprise. Much like the bottom-up philosophy, a process is instituted for requesting the addition of new applications.

Clearly each AW deployment philosophy has its strengths and weaknesses. Who owns, maintains and makes available the knowledge of which applications are '*good*' (also known as the enterprise application portfolio) across a large-scale network executing a huge variety of missions of varying opacity? As the size, diversity, and dynamism of an enterprise grows, it is likely that no one does. On the other hand, the bottom-up approach may add virus and other malware to the whitelist. For example, if an end host has previously been compromised with some form of malware, then that malware will be added to that end hosts whitelist and allowed to continue to run. This may be limited to that end host or, if the deployment strategy includes creating a whitelist superset that is then pushed out to all end hosts, may be allowed to run across the enterprise. Like many cyber defense technologies, deployment and employment of AW requires a strategic approach to achieve success; it is not a 'plug-and-play' solution.

A way to gain an intuition about the deployment of AW and how we approach the problem is to think of it as a low-dimensional solution-space, consisting of: the probability of disruption to business / mission operations, the probability of remediation of existing threats / ongoing exploits, and the probability of future exploitation after solution deployment. See Table 1 for a comparison of these qualitative dimensions of the problem at hand.

**Table 1 – Problem space**

| Dimension of Concern | Bottom-up | Top-down |
|---|---|---|
| Disruption to operations | Virtually none | Extremely high |
| Remediation of existing threats | Virtually none | Extremely high |
| Prevention of future exploitation | Extremely high | Extremely high |

The impact of the dynamism of the enterprise's or agency's mission or environment cannot be overstressed. For example, following an AW deployment of "bottom-up", one would not expect there to be a very large surge in AW requests for new software to be included on the whitelist which would require an increase in manpower to handle. While it may be expected that the bottom-up deployment approach would alleviate this, a dynamic environment or mission set could offset any expected benefit.

# Cyber Defense as a Complex Adaptive System

Our hypothetical government agency is a complex adaptive system, as discussed below. However, it does not end there. The cyber defense environment within which the government agency is situated is also a complex adaptive system. This one is made up of cyber attackers and defenders adapting to each other over time in a dynamic environment. AW may have a significant impact on the dynamics of this complex adaptive system.

## Complex Adaptive Systems

> *"When a scientist faces a complicated world, traditional tools that rely on reducing the system to its atomic elements allow us to gain insight. Unfortunately, using these same tools to understand complex worlds fails, because it becomes impossible to reduce the system without killing it. The ability to collect and pin to a board all of the insects that live in the garden does little to lend insight into the ecosystem contained therein"* (Miller and Page 2007).

A complex adaptive system is one made up of interacting components (agents) that adapt their behavior overtime in reaction to changes with respect to their environment and to each other (Holland 1995). The many interacting people and machines across our government agency constitutes a complex adaptive system (Miller and Page 2007) which must make ongoing local adaptions within a dynamic environment to successfully perform both its micro- and macro-scale functions. In fact, one could view the government agency enterprise as a 'superorganism' composed of a collection of individual organisms with division of labor/activities akin to those of a single organism (maintenance, human resources, research and development, budgeting, and so forth) (Hölldobler and Wilson 2009). Diversity in the system is driven role within the agency and by the interactions among the subsystems and the dynamics of their local microenvironment. Here, the local microenvironment is defined as an end host/user pair and the categories of software required to accomplish a mission. There is variation within the categories and bias from a social network to employ certain applications. There is also pressure from a dynamic environment, where missions and the application types required to accomplish them, change over time.

## Cyber Defense at our Hypothetical Government Agency

Our hypothetical government agency has, for years, used numerous firewalls and traditional AV software to protect its networks. Unfortunately, the persistence and sophistication of external hackers has been increasing rapidly over the past few years and defense has become increasing difficult, especially when premised on a reactive posture. This, coupled with shrinking budgets for cyber defense staff, has led the agency to consider AW as a way to make the agency's networks more secure without necessitating increased personnel budgets. The vast majority of the government agency's end host cyber defense capabilities are operated, managed and administered from two locations. These are known as Centralized Network Operations (CNO)-A and CNO-B. Any AW solution will be configured and deployed to the 1M+ end hosts by the operational personnel that comprise these two centralized network control centers. Any AW deployment strategy must be carefully considered since minor changes to the initial conditions of deployment may have a tremendous impact on the resulting dynamics and capabilities of the agency, such as the behavior of the enterprise with respect to generating and the CNOs' handling of change requests. These dynamics are greatly affected by the architecture of the technical solution, as discussed above. Each type of whitelisting solution will, by default, inherit the organizational architecture of the enterprise. That means that employment of either of these methodologies without addressing the architectural context may be very disruptive to the critical dimension under consideration: ongoing business/missions. For our purposes within this paper, we assume that large enterprises, such as our hypothetical government agency, have an extremely large and varied collection of *business/mission critical* applications of which there is no centrally held list. Clearly, this assumption is not valid for all large organizations, for example a telemarking firm, but those cases are being set aside for the current discussion.

In static environments with homogeneous end hosts, this centralization is inconsequential to the abilities of the agency. We hypothesize that in dynamic environments and/or ones with changing missions/priorities, this centralization may become a major limiting factor in successful AW employment across the enterprise; thus, necessitating a more decentralized management strategy (Anderson 2004).

### Top-down Whitelist Creation

To apply a whitelist across the enterprise is to impose homogeneity across the enterprise. While this will result in a potentially more secure set of cyber networks, the resulting in a superorganism is potentially less adaptive, less fit, and less able to accomplish dynamic missions within a dynamic environment as the ability to adapt the software in use is, potentially, much slower and, as such, may become misaligned with the temporal dynamics of the environment.

### Bottom-up Whitelist Creation

While the bottom-up strategy maintains the initial diversity of the agency, as discussed above, it may also maintain the initial set of malware resident on an agency's networks. While traditional AV software could be used to eliminate *known* threats zero-day exploits would still be a threat. Furthermore, the bottom-up strategy may still produce an agency poorly able to deal with a highly dynamic environment.

More generally, the unintended consequences of centralized decision making, an inevitable outcome of AW in top-down deployments, include the possibility of breaking missions that are currently underway; this could potentially involve risk of loss of life. Another unintended consequence of a centrally controlled, top-down application whitelisting scenario is the possibility of overwhelming the CNOs with application approval requests for which they are not prepared to handle. The bottleneck that may be created could also pose a great hazard to operational continuity at some point in the future, well *after* deployment. Unintended consequences on highly centralized "bottom-up" deployments are similar, but thought not to be as problematic.

As highlighted in the above discussion, the dynamics of AW are surprising complicated and coupled to many other systems. Given the adaptive and dynamic nature of the components of the system, traditional, reductionist, methods of analysis may miss important aspects of the system as it is the interaction among system components that generate many of the elements of interest. In the sections that follow we define an analytic process based upon complexity science that can be used to analyze such a system.

## Methods

While there are many tools available for analyzing the security posture of end hosts, such as metasploit (https://www.metasploit.com), these tools assume a static environment and do not include the impact of the behaviors of the users. As such, these tools may miss a very important driver of the dynamics of these systems. With modern software and hardware it is now possible to explicitly represent all of the components of a system and simulate the systems dynamics over time. When studied in this manner one can gain a much better understanding of system dynamics (Axtell 2000). Moreover, this is, except in trivial cases, the most efficient way to understand the temporal dynamics of a system and its future states (Buss, et al. 1991).In what follows we will define a minimal agent-based model (Axtell 2000) that can be used to begin to study the AW system. An agent-based model (ABM) is a simulation tool that explicitly represents the components of a system *and* how these components interact with each other and change over time (Axtell 2000).

Complexity science, and here, specifically agent-based modeling, supplies us with an evidence-based method of weighing options beyond conjecture and intuition. Tools for dealing with evolutionary and adaptive dynamics are not commonplace in operations research methodologies. The focus here is on the end host's relationship to the mission environment as well as the host's relationship to other hosts in a social network, which biases a host's knowledge of available applications and their choice of which application to use, as well as on the dynamism to which the end host will be subjected. That dynamic, in turn, is affected by the use of AW.

### The Whitelisting Model Formulation

As discussed above, the personnel within our hypothetical government agency are trying to accomplish their mission(s) with various sets of software. The quicker the person or agency can figure out what that software should be, the quicker it can respond to changes in environment. This is the basic dynamic depicted in the agent-based model we created to study this system. Moreover, the system-wide goal of the model is for whitelisting to cause as little disruption as possible to the adaptation abilities of the hosts (a collection of heterogeneous subsystems that comprise the agency). The simulation was created using the NetLogo agent-based modeling framework (Wilensky 1999).

### Model Overview

The agents primarily exist in an abstracted network space. There are two networks at play among the agents. The first network is the information technology network made up of the two CNOs and the end hosts connected to them. The second network is a social network, over which agent exchange information about software that can be used to accomplish a particular task or mission. Secondarily, agents exist within an abstract non-torodal 2-deminsional

"problem space." This problem space defines the problems faced by the agents which, in turn, defines the agents' software needs. Over this dynamic is the AW deployment. In the simulation this can take two forms, as described above, top-down with a single approved AW; or bottom-up, with each end host creating its own initial AW. As time progresses, the agents move around the problem space which means they may need a different set of software applications to accomplish their mission. Due to the AW security posture, any changes to an end host's software set requires approval by the CNO to which they are connected. Each request is added to a request queue at the CNO and, eventually, processed. When processed the software request may or may not be approved. At each time step of the simulation data on request queue length, waiting times, and mismatches between an end hosts software list and what is needed in their problem space are gathered.

## Model Details

First the problem space is created. This problem space comprises a 401 x 401 grid of 160,801 patches. Each patch contains a bit string of user defined length. This bit string represents the types of software needed to achieve a mission on the patch. It should be noted, and will be discussed below, that this bit string only defines the types of software needed, not the specific software solution to use. For example, a bit may say a spreadsheet is needed, but will not define which spreadsheet software an end host should use.

The agents populating the model are created via a set of parameters. The first is the number of agency networks. The second is the number of CNOs per agency, and finally, the number of end hosts per CNO. In order to study a single agency, we choose to fix the number of agency networks at one. Moreover, in order to better emulate current agency operations, we choose to hold the number of CNOs fixed at two.

Upon instantiation end hosts are provided with a bit string the same length as the bit string of the patches. The difference, however, is that the end hosts' bit string contains integer values from 0 to 9, rather than only 0 and 1 as does the patches' bit string. The increased diversity was used to emulate the fact that numerous software solutions exist for any given functionality need, i.e., there are many spreadsheet programs available. This is the information that is exchanged among agents via their social network. When an end host discovers a need for a new software application (when the bit string of the patch the end host is on contains a 1 and the end host's bit string in the same position contains a 0), the end host queries its social network for options, accepting the majority's opinion. If no one in their social network is using an application for that task, then the end host chooses a software application at random (randomly choosing an integer from 1 to 9).

This dynamic is allowed to proceed unimpeded for an initial 'burn in' period; thus, allowing the hypothetical government agency to evolve to the needs of its environment. After the burn in period, the AW security posture is rolled out. Now, when an end host discovers a discrepancy between its list of available software applications and the environmental needs, it must submit a request for approval to its CNO. This approval will require some period of time (log-normally distributed) to complete and may or may not result in approval. If the requested software has been approved previously, the process takes less time and, with certainty, ends in approval. The overall system flow is shown in Figures 1 and 2.

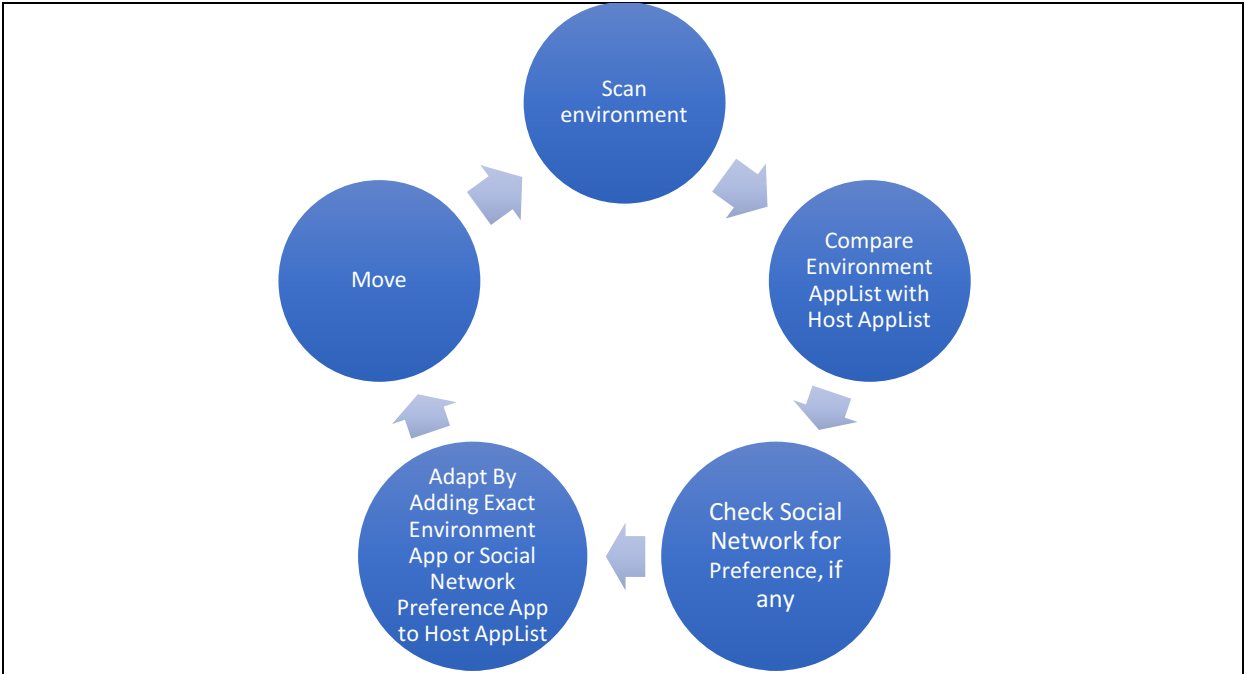

Figure 1 - High level logical flow prior to AW deployment

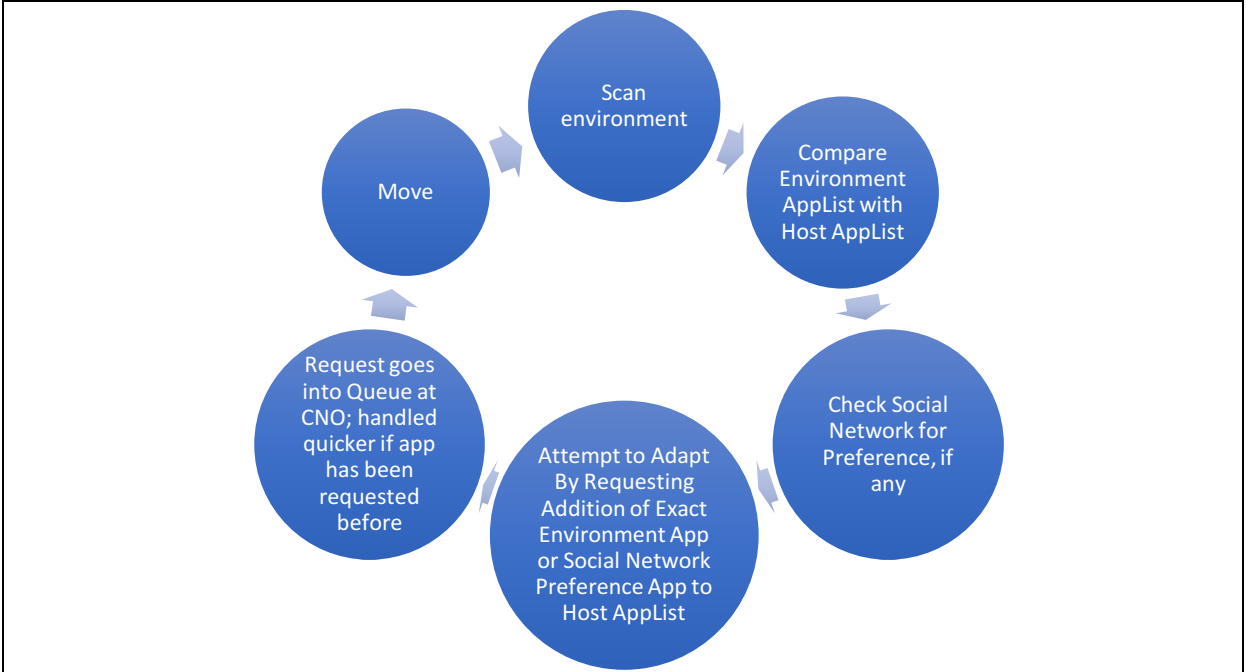

Figure 2 - High level logical flow after AW deployment

The CNOs are made up of a set of analysts that process application approval requests. The more analysts that are part of the CNO the faster requests can be processed. Furthermore, the CNO, as discussed above, after the burn in period may roll out the AW in two ways: 1) top-down or 2) bottom-up. The simulation is run for 2500 time steps. Model dynamics were loosely tailored to correlate a time step with a day; therefore, each simulation was run for approximately 7 years of simulated time.

### The Design of Experiments

In order to test the dynamics of this system under varying assumptions, we defined the design of experiments (DOE) shown in Table 2. This DOE was created to test the impact of: AW roll out via top-down or bottom-up, the heterogeneity and dynamism of the environment, size of the application lists (potential needs), the efficiency of request processing, and the number of end hosts. This results in a DOE made up of nine parameters each comprising a high/low value, yielding 512 discrete design points. Each design point was run for 12 replications (made up of 12 random seeds held constant across all design points. This created 6144 individual runs of the model.

Intuitively, it would seem that applying a whitelist tailored to the end host would solve the problem of request surge upon deployment, but our intuition is that this only holds for a static environment. We also expect that higher throughput for requests at the CNO will help alleviate the potential bottlenecks associated with the adoption of new applications caused by AW.

Table 2 – DOE

| Parameter | Note | Values |
| --- | --- | --- |
| isMissionSafe | Bottom up / Top down | True, False |
| sizeAppList | Length of application portfolio | 10, 100 |
| dynamicEnv | Dynamic environment | True, False |
| dyProb | Probability of a change in the local environment (%) | 25, 75 |
| arePatchesSame | Homogeneity of environmental initial conditions | True, False |
| processProbability | Request processing efficiency (%) | 25, 75 |
| numLeaves | Number of end hosts per server | 100, 750 |

### Results

For the most part, the results correspond to our aforementioned intuition about model dynamics. Overall, AW decreased our hypothetical government agency's ability to adapt to a dynamic environment. As the number of end hosts per CNO decreases or the capacity of the CNOs to process requests increases, the impact of the AW on agency adaptation can be mitigated. Our main response variable was the average length of request queues at the CNOs. Over all runs, the average length of request queues at both CNOs is shown in Figures 3 and 4. As can be seen in Figures 3 and 4, there are two basic dynamics, one in which there is an initial surge in requests, and subsequent increase in request queue length upon roll out of the AW, followed by a decrease in queue length. The other basic dynamic is a simple, near monotonic increase in queue length after the AW roll out. These two regimes correspond to the interplay between demand for application approval, driven by the heterogeneity and dynamism of the environment and size of the end host population, and the CNO practices, which are a function of AW roll out strategy and application approval request processing efficiency. Interestingly, the dynamism of the environment does not appear to be a significant driver of system dynamics, as demonstrated by the only minor differences in Figures 3 and 4.

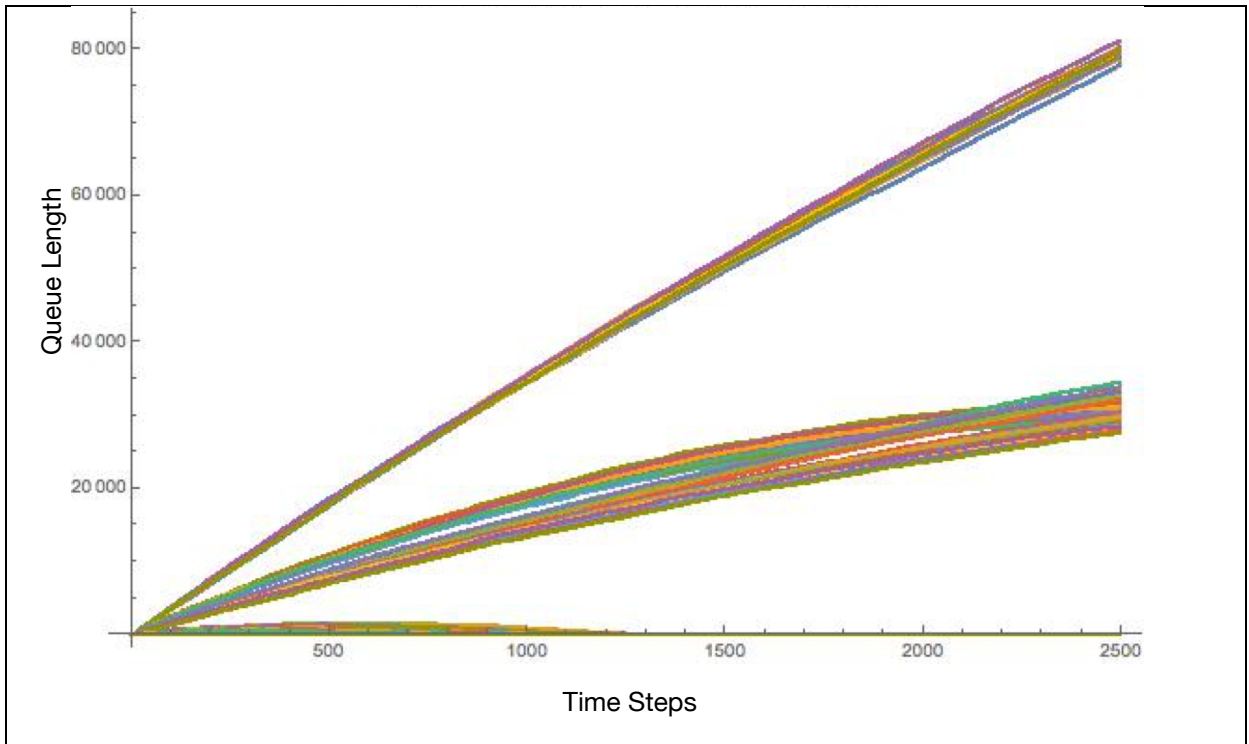

Figure 3. Average request queue length at the CNOs over time with a dynamic environment.

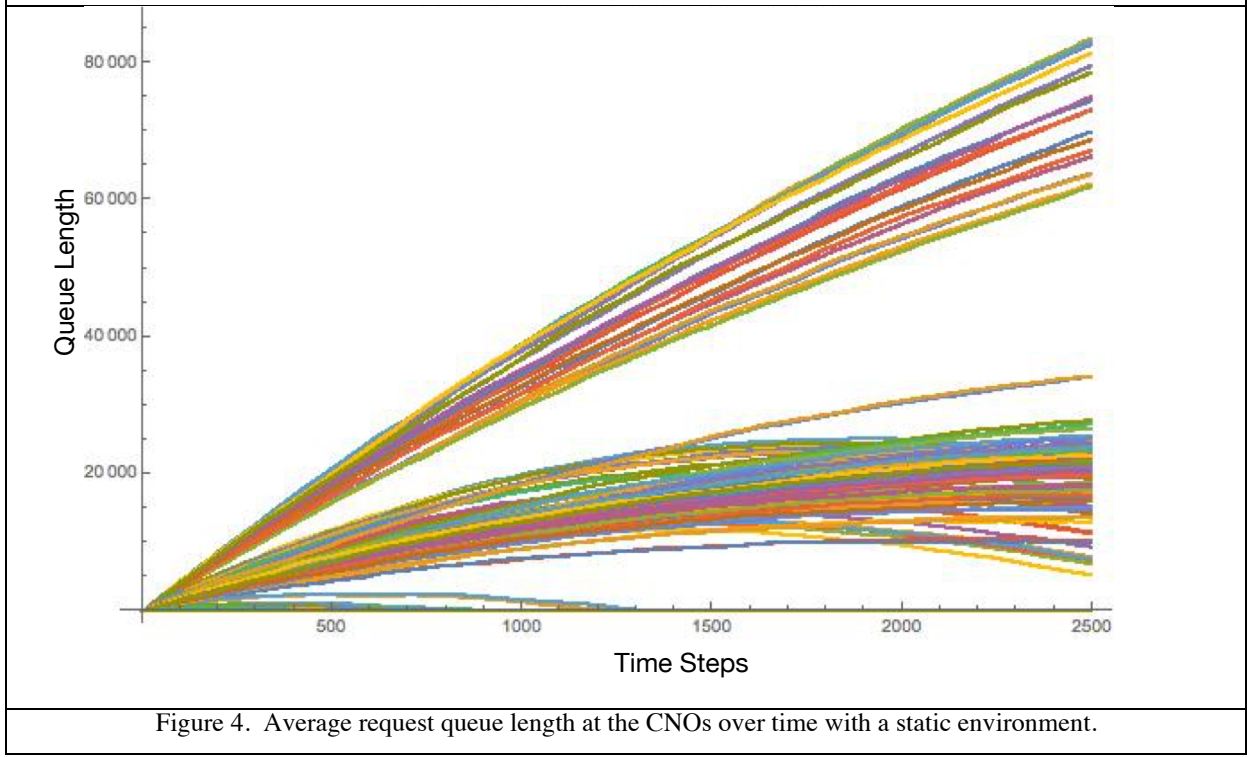

Figure 4. Average request queue length at the CNOs over time with a static environment.

As shown in Figures 5 and 6, the efficiency of queue processing and the size of the end host population were significant drivers of request queue length dynamics.

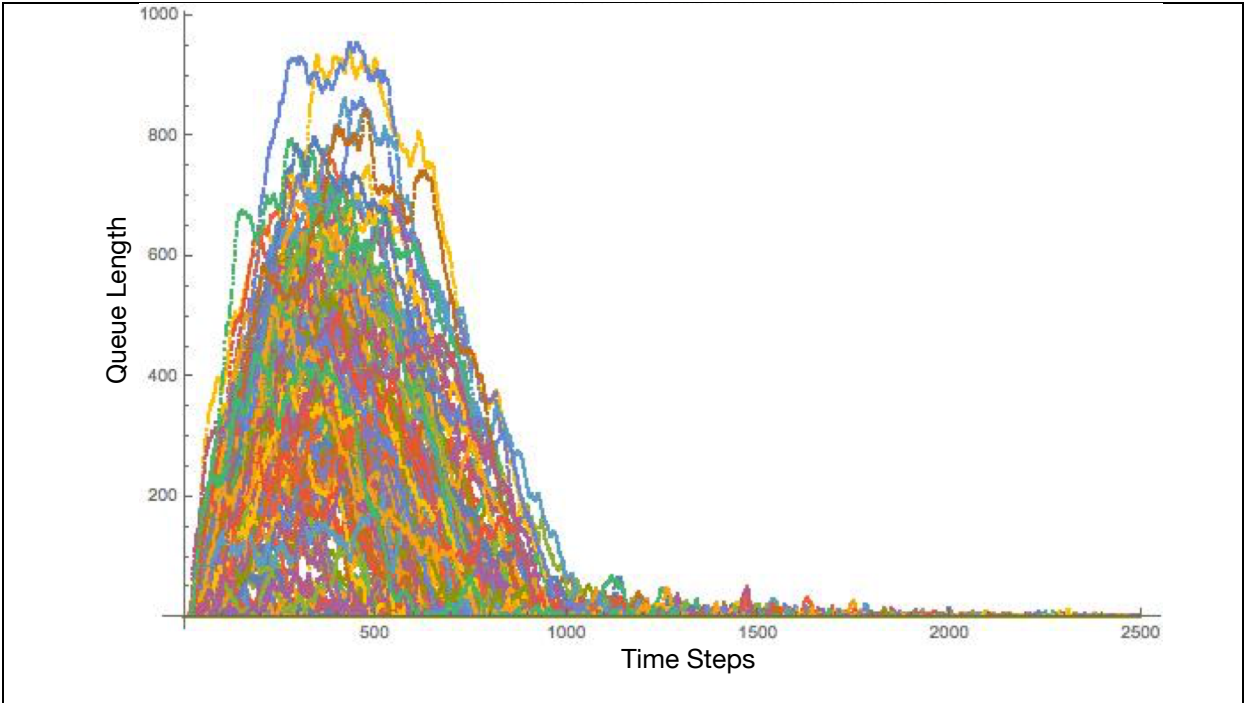

Figure 5. Application request queue length for all model runs over 2500 days that had application/environment strings of length 100, 75% queue processing efficiency and employed bottom-up whitelist creation.

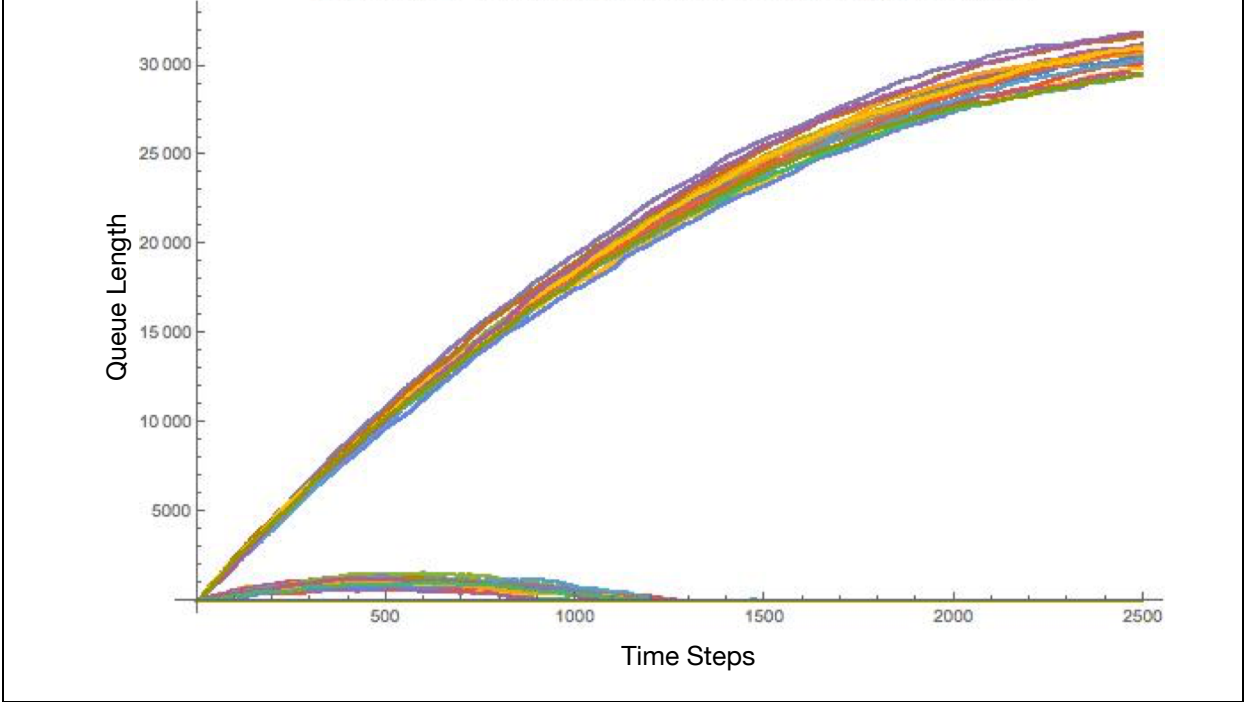

Figure 6. Application request queue length for all model runs over 2500 days that had application/environment strings of length 100, 75% queue processing efficiency, had a dynamic environment, and employed top-down whitelist creation.

# Conclusion and Next Steps

Clearly the model discussed herein is only a first step in the analysis of the dynamics associated with the implementation and long term use of AW. We feel that this model represents a reasonable foundation for studies of AW. However, even a model of this simplicity brings up interesting and counterintuitive results. For example, the use of top-down vs. bottom-up roll out of the AW had a less significant impact over time than we expected. The most significant driver appears to be how quickly CNOs can process requests. This implies that CNOs will need to be very well staffed, at least initially, in order to implement a minimally disruptive AW roll out. This, of course, would be extremely expensive. Alternatively, one may attempt to implement some sort of automated system for application approval. However, this may suffer from logic that fails to approve necessary applications or approves unnecessary applications that introduce vulnerabilities into the agency. Finally, regardless of how an agency rolls out its AW, it should be expected that there will be an immediate, potentially very large, surge in applications for approval; this is especially true if the agency is in a dynamic environment or has a dynamic set of missions.

We now have the foundation in place to take a large leap forward in our ability to model cyber defense strategies. We can use the flexibility of the platform to explore alternative architectural configurations to inform the agency's employment strategy. How can we alter the system's architecture or rules to get more predictive behavior when in a dynamic context? For the case of our AW model for a government agency, could one implement a prioritized queue so that critical agency missions' requests are reviewed immediately at the expense of less critical missions?

Currently, the model has relational equivalence (Axtell 1996) to the real world, which means it behaves in a manner that is logically consistent with the real world; basically, more application approval requests equate to longer queues subject to CNO processing efficiency. While this may be appropriate for this initial foundation, as the simulation grows in detail, we also plan to collect data to inform model assumptions and parameters and preform a more formal validation. A formal validation would allow us to continue to build faith in the utility of surprising or counterintuitive results.

There are many directions in which to take continued development. For example, operational data such as the frequency of new applications requests, as well as stratification information (i.e., who makes a lot of requests and how many 'heavy requesters' are there?) and social network information (i.e., what is the makeup of the social network within the enterprise?). Simplifying assumptions have been made about some of these parameters. These assumptions were made conservatively and can be tuned or parameterized as data is collected. If, for example, the real world social network, likely, is much more densely connected than the one we have modeled, this, in turn, may change the rate of application adoption. While we don't anticipate that switching application preferences will have a major impact on the queue dynamics, it would be preferred that data informs those model parameters. The model is also built upon the assumption that all application options within a given category are equally effective at accomplishing a given mission. Since we are not asking questions about mission effectiveness at this time, this assumption is reasonable.

We are very interested in scaling the model up to see how that affects the results. Due to computational time constraints, most of our results come from a network of 200 end hosts (a subset of the agency was run with 750 end hosts). We believe scaling this up to an order of magnitude is much closer to the reality of the agency's network and is of paramount importance to ensure realistic model results.

Finally, we would like to incorporate the effects of malware on AW deployment and end host skills. If an undetected exploit existed in a bottom-up configuration, that exploit could hypothetically end up whitelisted. Along with the effects of malware, we feel it would be useful to incorporate the effects of heterogeneity of the technical expertise of leaf nodes; not all end users are of the same technical expertise level, and this may have an impact on how whitelists are managed. For example, a tech-savvy user may be given the ability to approve their own whitelist requests. This type of user would also be more likely to know when a program execution request is made that should not have been and be able to alert the appropriate response authority.